%% file: manuscript.tex
\documentclass[a4paper,fleqn]{cas-sc}
\usepackage{hyperref}

\usepackage[round]{natbib}
\usepackage{graphicx}
\usepackage{xspace}	
\usepackage{graphicx}
\usepackage{lscape}
\usepackage{threeparttable}
\usepackage{amsmath}
\usepackage[utf8]{inputenc}
\usepackage{longtable}
\usepackage{graphicx}
\usepackage{multirow}
\usepackage{rotating}

\begin{document}
\let\WriteBookmarks\relax
\def\floatpagepagefraction{1}
\def\textpagefraction{.001}

\shorttitle{Mapping reddening plane in the Galactic disk}

\shortauthors{Joshi, Y. C. 2025}

\title [mode = title]{Mapping the reddening plane in the Galactic disk through interstellar extinction of open clusters} 

\author[1]{Y. C. Joshi}[type=author,
                        auid=000,bioid=1,
                        orcid=0000-0002-4870-9436]

\cormark[1]

\ead{yogesh@aries.res.in}

\affiliation[1]{organization={Aryabhatta Research Institute of Observational Sciences},
    addressline={Manora Peak}, 
    city={Nainital},
    postcode={263001}, 
    country={India}}

\cortext[cor1]{Y. C. Joshi}

\begin{abstract}
As thousands of new open clusters in the Galaxy have recently been reported with reddening or extinction information, we map the distribution and properties of the MilkyWay’s interstellar material in the Galactic disk as traced by these open clusters. By analyzing the distribution of interstellar extinction for 6215 open clusters located at low Galactic latitude (\(\lvert b \rvert \leq 6^{\circ}\)), corresponding to the thin Galactic disk, we identify a reddening plane characterized by a dust layer whose thickness varies with Galactic longitude. By splitting the open clusters sample into several sub-regions of Galactic longitude, we observe that the reddening plane is not perfectly aligned with the formal Galactic plane, but instead varies sinusoidally around the Galactic mid-plane. The maximum and minimum interstellar absorption occur at approximately $42^{o}$ and $222^{o}$, respectively, along the Galactic longitude. Our analysis reveals a noticeable north-south asymmetry in the distribution of interstellar absorption, with a higher proportion of interstellar material below the Galactic plane. We also find that the Sun is located \(15.7 \pm 7.3\) pc above the reddening plane. The scale height of the open clusters from the reddening plane is estimated to be $z_h = 87.3 \pm 1.8$ pc. The mean thickness of the absorbing material in the reddening plane, which represents the average extent of the dust layer responsible for interstellar extinction, is found to be about $201 \pm 20$ pc. Our findings provide insights into the distribution of interstellar dust, its relationship with the Galactic thin disk, and its implications on the Galactic structure.
\end{abstract}

\begin{keywords}
ISM: dust, extinction \sep Galaxy: open clusters \sep general: methods: statistical \sep astronomical databases
\end{keywords}

\maketitle

\section{Introduction}\label{sec:intro}
Interstellar material, which includes dust and gas, can absorb and scatter light from celestial sources. This can make it more challenging to observe distant objects, as the intervening material can dim or redden the light, affecting our ability to study them accurately \citep{2017MNRAS.464.2545C, 2024MNRAS.527.4863U}. By studying the distribution of interstellar extinction along different lines of sight in the Galaxy, one can gain valuable insights into the structure and composition of the Galactic disk, particularly the thin disk, where most interstellar material is concentrated. This includes parameters such as the scale height which characterizes the vertical extent of the disk, and Solar offset which characterizes the distance of the Sun from a reference plane. Understanding these properties of the interstellar medium is crucial for accurately interpreting observations and inferring the true characteristics of celestial sources, especially those located at significant distances from us. Knowing the local distribution of dust also enhances our understanding of the overall structure and dynamics of our Galaxy as well as the probable sites of star formation \citep{2005MNRAS.362.1259J, 2014MNRAS.443.1192C, 2018A&A...618A.168R}. However, due to the obscuration caused by line-of-sight dust extinction in photometric observations, our knowledge of the low-latitude Galactic disk structure has been limited \citep{2024A&A...692A.255R}. In this context, open clusters, which are cohesive groups of large number of stars and intrinsically bright in nature, can serve as valuable tracers of the Galactic structure, particularly around the mid-plane.

Open clusters (OCs) are primarily located within the Galactic disk, and are visible out to large distances, with their distribution peaking at a Galactocentric distance of approximately 7-10 kpc \citep{2016A&A...593A.116J} . The density of OCs decreases both toward the Galactic center, where tidal forces are stronger, and toward the outer Galactic disk, where star formation rate drops significantly. They exhibit a relatively thin vertical distribution compared to other Galactic components, with most clusters confined within $\pm$200 pc of the Galactic mid-plane ($z=0$). Since OCs form within dense molecular clouds, they are excellent sources to probe interstellar matter in the Galactic disk due to their proximity to the Galactic plane and their wide distribution across various Galactocentric distances. Furthermore, their reddening and extinction values can be precisely ascertained through multi-band photometric observations \citep{1987MNRAS.228..483S, 2007MNRAS.380.1141S}.

The use of OCs in studies of the distribution of interstellar matter is a valuable and time-tested idea. Numerous studies have been undertaken in the past using OCs to refine our models of the interstellar medium and its distribution in the Galactic plane \citep{1987MNRAS.226..635P, 1992A&A...258..104A, 1998A&A...336..137C, 2005MNRAS.362.1259J, 2007MNRAS.380.1141S, 2014MNRAS.443.1192C, 2021ApJ...906...47G}. In recent years, information on reddening, extinction, age, distance, and chemical composition has become available for several thousand OCs \citep{2023A&A...673A.114H, 2023AJ....166..170J, 2024FrASS..1148321J}. It is therefore imperative to investigate the distribution of reddening material in the Galactic disk and reassess correlations of interstellar material with different Galactic parameters. Although numerous 2D and 3D extinction maps of the sky have been produced \citep[e.g.,][]{2009ApJ...696..484B, 2014MNRAS.443.2907S, 2014ApJ...789...15S, 2019ApJ...887...93G, 2020A&A...641A..79H, 2025arXiv250302657Z}, few studies have focused on  the large-scale variation of extinction or interstellar dust in the Galactic plane using OCs \citep{1987MNRAS.226..635P, 2005MNRAS.362.1259J}. In this paper, we use the largest available sample of OCs with extinction and distance information to investigate the large-scale distribution of interstellar material in the Galaxy, emphasizing its concentration within the Galactic disk and examining its distribution in both radial and vertical directions.

This paper is structured as follows: Section~\ref{data} describes the data used in the present study.  Section~\ref{reddening} presents the analysis of Galactic extinction. In Section~\ref{galactic_absorption}, we examine the distribution of Galactic absorption across various longitudinal directions. Section~\ref{scale_height} discusses the determination of the cluster scale height relative to the reddening plane. Our findings are summarized in Section~\ref{summary}.
%
%------------------------ Fig. 01 --------------------------------------
\begin{figure*}[b]
\centering
\includegraphics[scale=0.38]{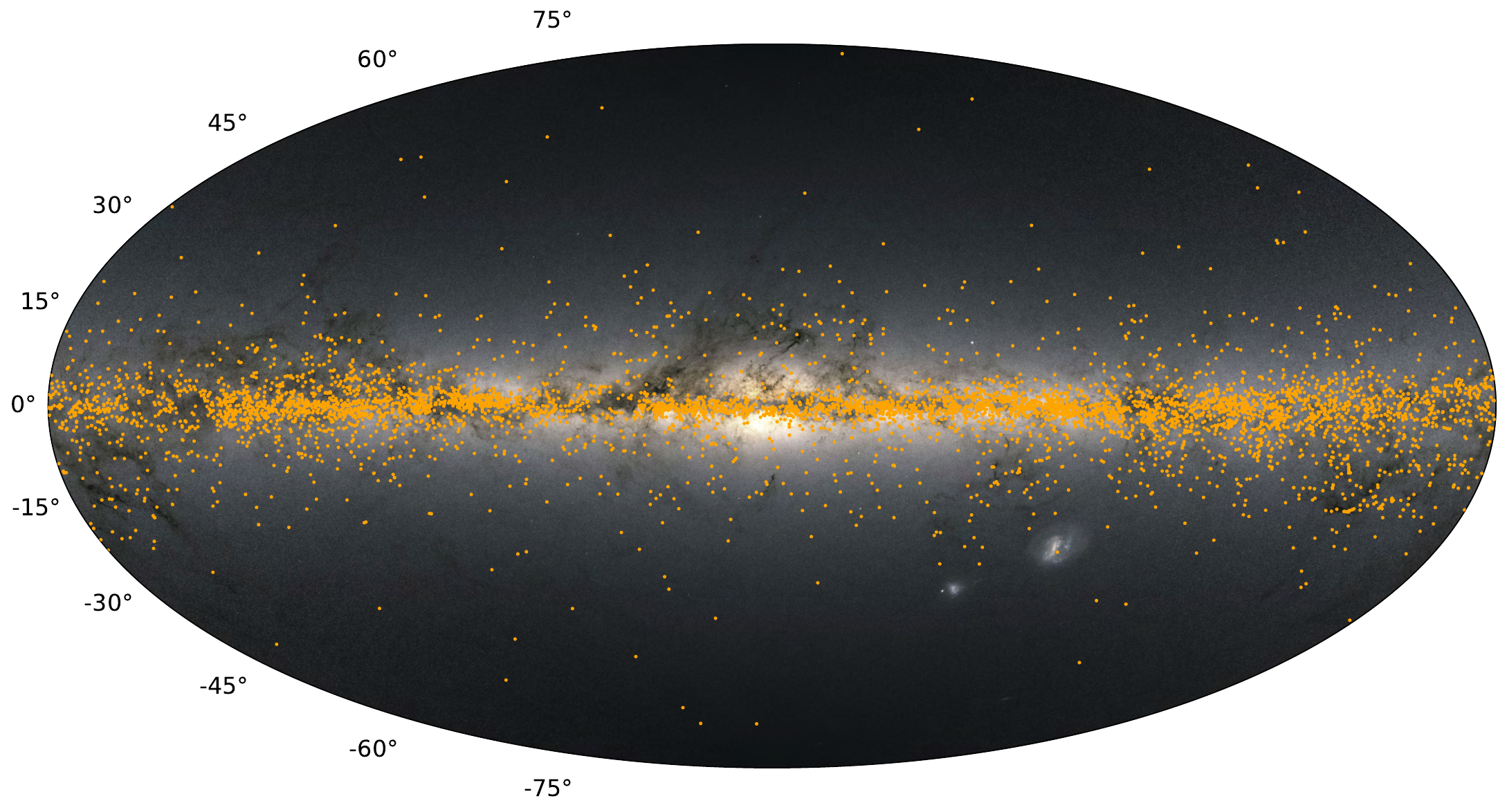}
\caption{An on-sky view of Milky Way  in the Galactic longitude-latitude ($l-b$) plane generated by ESA Gaia Early Data Release 3 (EDR3) juxtaposed with our sample of 6215 OCs. Most of the clusters are located in the Galactic mid-plane $b = 0^0$.}
\label{fig01}
\end{figure*}
%--------------------------------------------------------------
%
\section{Archival Data}\label{data}
To obtain a homogeneous sampling of the underlying OC population, a combined catalog was constructed out of several large-scale studies that provide physical parameters for a significant number of OCs. A detailed procedure for data compilation, including the removal of duplicate and spurious clusters, is extensively explained in \citet{2023AJ....166..170J}. Most of these clusters were identified in the Gaia era, with their parameters derived using machine learning tools due to the humongous volume of the data \citep{2018AA...618A..93C, 2018ApJ...856..152R, 2019A&A...624A.126C, 2019ApJS..245...32L, 2021MNRAS.504..356D, 2021RAA....21...93H, 2022AA...660A...4H, 2022AA...661A.118C, 2024NewAR..9901696C, 2024arXiv241015305R}. With accurate estimates of extinction along various lines of sight and precise distance measurements now available for these open clusters, it is possible to better understand the distribution of interstellar material in the Galactic disk and probe the Galactic structure.

In the present work, we retrieved a catalog comprising 6215 OCs from these surveys, and parameters such as age, reddening, extinction, and distance estimates have been compiled. For the clusters, where only color excess or reddening $E(B-V)$ was provided, we calculated the line-of-sight dust extinction as$A_V = R_V \times E(B-V)$ for each cluster. Here, we used a standard total-to-selective extinction ratio of  \(R_V\) as 3.1 \citep{1989ApJ...345..245C}. We excluded some recent OC catalogs that do not provide extinction or reddening estimates for the newly identified clusters. Most of the extinction values for clusters were derived from photometric observations. Figure~\ref{fig01} illustrates an on-sky view of the Galaxy in the Galactic longitude-latitude ($l-b$) plane, with the current sample of OCs superimposed. Since some recent studies \citep[e.g.,][]{2022ApJS..260....8H, 2023ApJS..265...12Q} provided reddening estimates in the G band, we used the relation given by \cite{2018MNRAS.479L.102C}  [\(A_G = 2.74\cdot E(B-V)\)] to convert \(A_G\) to \(A_V\). However, due to the unavailability of uncertainty estimates in the  \(A_V\) values for all open clusters, a weighted analysis could not be performed in the present study.
\section{Galactic distribution of the reddening material}\label{reddening}
The spatial distribution of reddening material provides vital clues about star formation through the interplay between dust and gas distribution. Studying interstellar extinction, $A_V$, is essential for tracing these distributions and offers insights into the structure and evolution of the Galaxy \citep{1987MNRAS.226..635P, 1998A&A...336..137C, 2014MNRAS.443.2907S, 2020A&A...641A..79H}. The patchy distribution of dust and gas within the Galactic disk poses challenges in accurately determining $A_V$ values, especially for young OCs. The present catalog of 6215 OCs thus enables a comprehensive investigation of the Galactic reddening distribution. 

\begin{figure}[h]
\centering
\includegraphics[scale=0.35]{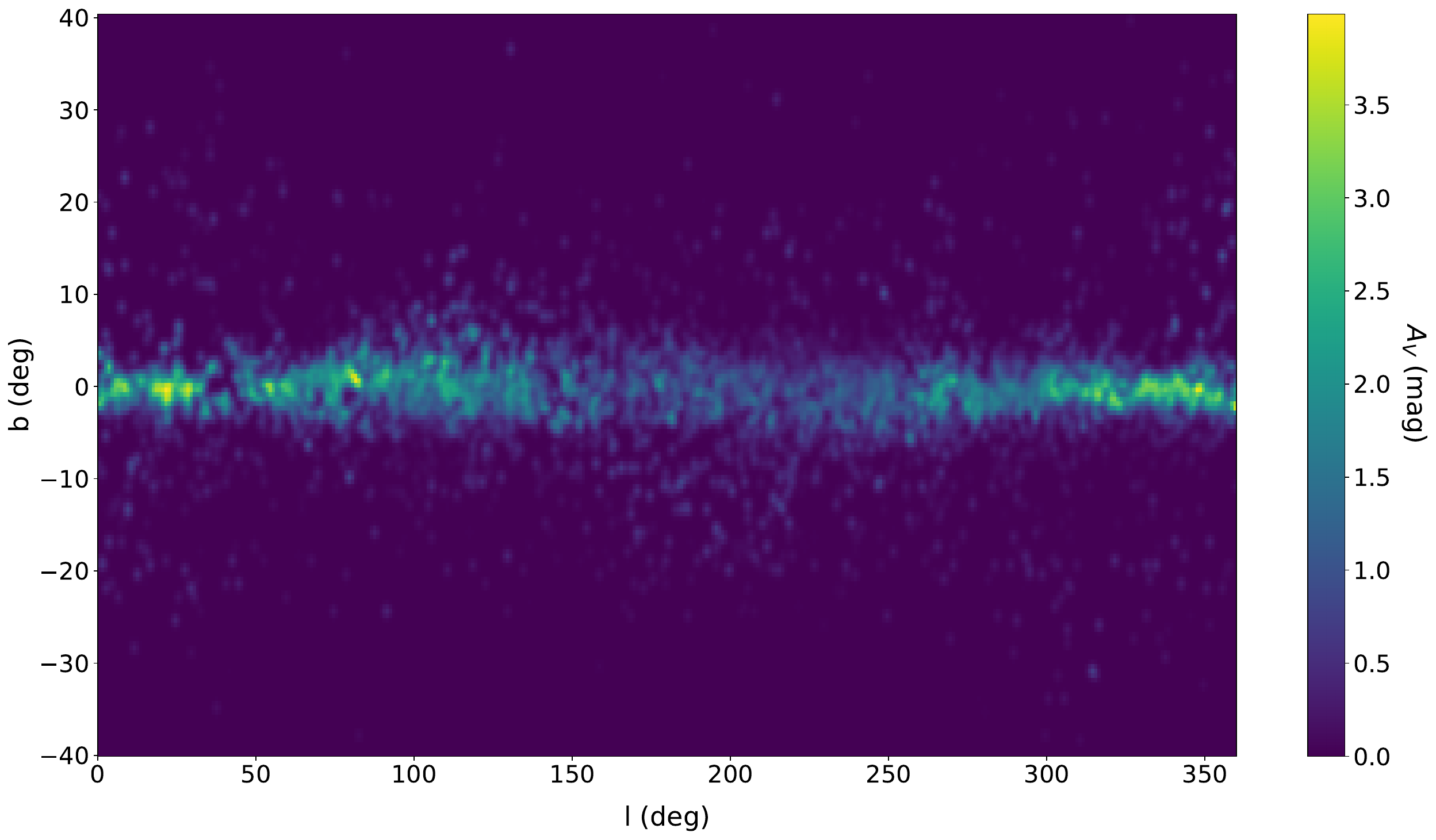}
\caption{Heat map of the distribution of reddening material in the Galactic $l-b$ plane. The color in the map represents the mean extinction of the cluster.}
\label{fig02}
\end{figure}

In Figure~\ref{fig02}, we present a heat map of the reddening material distribution in the $l-b$ plane, where color indicates the amount of reddening at the cluster positions. Several dust clumps are clearly visible in the figure along the Galactic plane. Of our dataset, 5041 OCs ($\sim 80\%$) are located within \(\lvert b \rvert = 6^{\circ}\). Given the concentration of reddening material near the Galactic mid-plane, where most interstellar matter resides, we restrict our sample to this region (\(\lvert b \rvert = 6^{\circ}\)) for all subsequent analyses, focusing exclusively on the thin disk structure.
%
%--------------------------- Fig. 02 -------------------------
%----------------------------------------------------------------
%
\subsection{Vertical distribution in the Galactic plane}\label{gp_p}
\begin{figure}[b]
\centering
\includegraphics[scale=0.35]{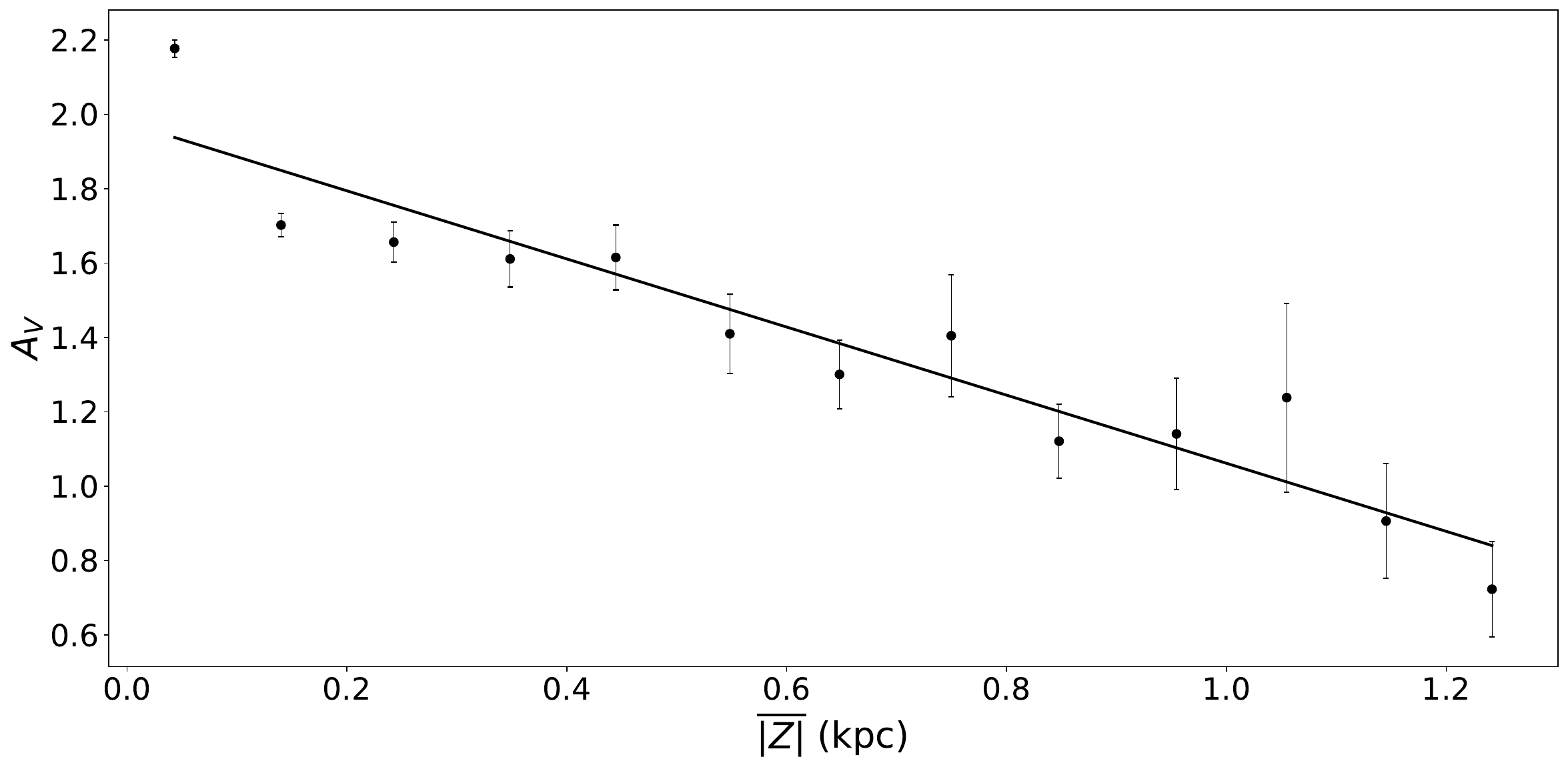}
\caption{Mean value of \(A_V\) as a function of mean \(\lvert z \rvert\) for the same data sample. The bin width in \(\lvert z \rvert\) is fixed as 100 pc. The least squares fit is shown by a continuous line.}
\label{fig03}
\end{figure}

In our study of Galactic distribution of OCs carried out in \citet{2023AJ....166..170J}, we found that the majority of OCs are located below the Galactic plane. Since young OCs trace dust due to their formation and evolution within embedded star-forming regions, we expect the reddening material to also be asymmetrically distributed around the Galactic mid-plane. To further examine the distribution of interstellar extinction material in the vertical direction of the Galactic plane, we plotted the mean extinction in the direction of OCs as a function of their mean absolute vertical distance $z$, as shown in Figure~\ref{fig03}. Since the mean extinction gradually decreases along the Galactic disk, we fitted the distribution with using linear regression. The least-square fit for mean \(\lvert z \rvert - A_V\) yields the following relation
\begin{equation}
\hspace{5.0cm} A_V = -0.92 ~\overline{\lvert z \rvert} + 1.98
\end{equation}
\vspace{-0.5cm}
$$\pm 0.1 ~~~~~~~~~~\pm 0.07$$

The best fit is shown by a continuous line in Figure~\ref{fig03}. We also performed independent linear fits above and below the Galactic mid-plane for the mean reddening versus absolute vertical distance, and found the following relations.
\begin{equation}
\hspace{5.2cm} A_V = -0.90 ~\overline{z} + 1.99 \hspace{5mm} (z >= 0)
\end{equation}
\vspace{-0.5cm}
$$\pm 0.11 ~~~\pm 0.08$$
\vspace{-0.5cm}
\begin{equation}
\hspace{5.2cm} A_V = -0.94 ~\overline{z} + 1.96 \hspace{5mm} (z < 0)
\end{equation}
\vspace{-0.5cm}
$$\pm 0.13 ~~~\pm 0.09$$
%
%--------------------------- Fig. 03 ---------------------------------
%--------------------------------------------------------------------- 
%
The extinction exhibits a similar variation in  both the northern and southern Galactic hemispheres, indicating a consistent extinction pattern in both directions. The typical slope of $\frac{dA_V}{dz}$ is determined to be $-0.9\pm0.1$ mag\,pc$^{-1}$, although extinction remains relatively unchanged beyond 1 kpc in the z-direction. This negative slope indicates a continuous decrease in the amount of interstellar reddening material as one moves away from the Galactic mid-plane. This result aligns with expectations, as older clusters, typically located farther from the Galactic plane, contain little dust or gas, with most of the material either consumed in star formation or dispersed over time. In our earlier study, based on the reddening information for a relatively smaller sample of 1211 OCs within 1.8 kpc of the Sun, we found a reddening slope ($E(B-V)$ vs \(\lvert z \rvert\)) as $0.41\pm0.05$ mag\,pc$^{-1}$ \citep{2016A&A...593A.116J}. Given $R_V=3.1$, it implies a slope of $\frac{dA_V}{dz} \sim 1.27$, indicating that the extinction gradient in the current study is slightly flatter than in the previous one.
\subsection{Distribution along the Galactic plane}\label{gp_p}
%
%--------------------------- Fig. 04 -----------------------------------
\begin{figure}[b]
\centering
\includegraphics[scale=0.35]{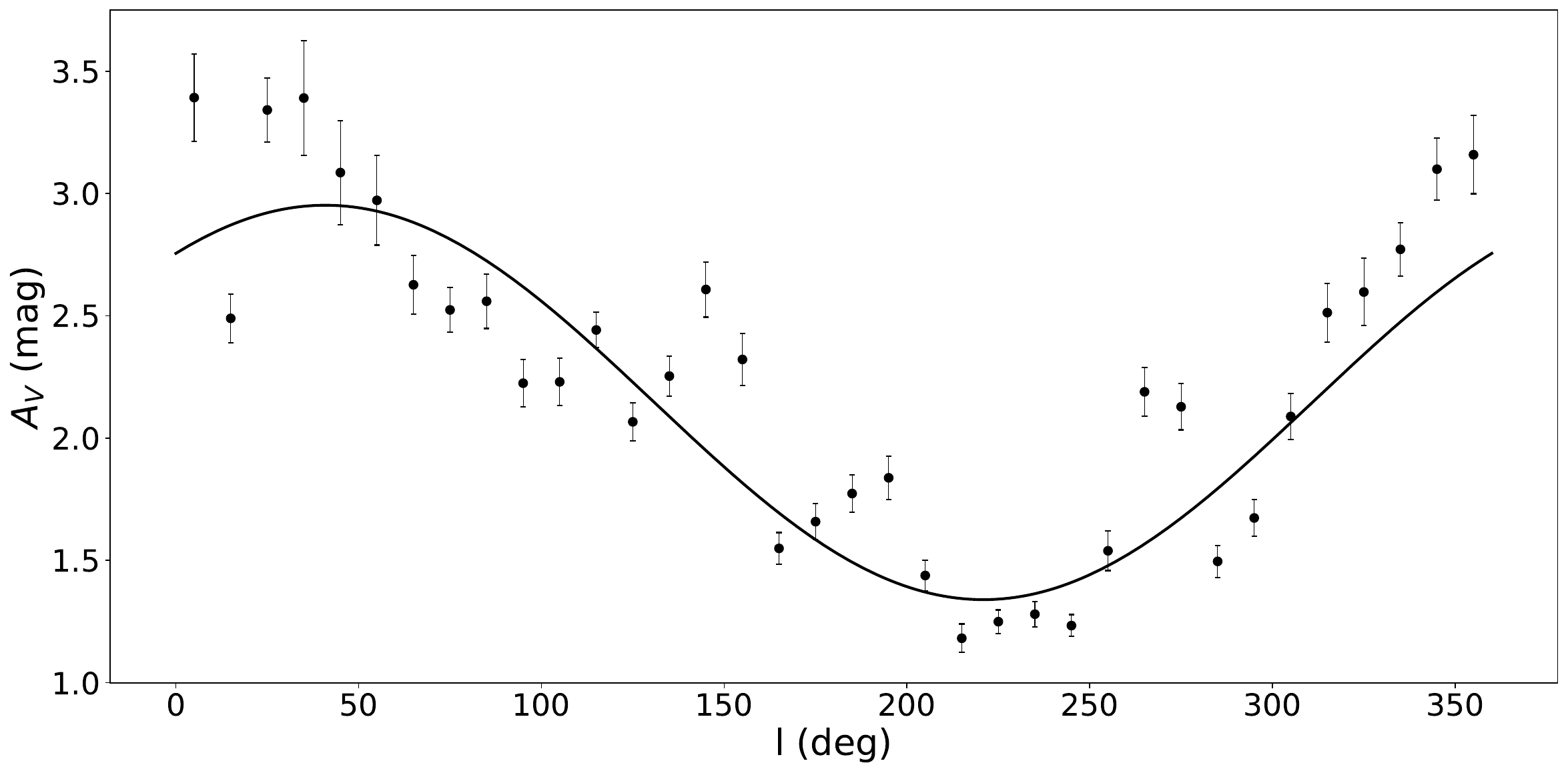}
\caption{Mean extinction as a function of Galactic longitude in bins of \(10^{o}\) along with the sinusoidal fit depicted by a continuous line.}
\label{fig04}
\end{figure}
%--------------------------------------------------------------
%
Figure~\ref{fig04} illustrates the variation of mean extinction in \(10^{o}\) bins of Galactic longitude. The plot reveals greater scatter in mean extinction towards the Galactic center, indicating significant variability in the amount of interstellar material in that direction.  In contrast, the anti-center region exhibits less scatter and more systematic variations in extinction. Although clusters at the same Galactic longitude show a range of reddening values due to their varying positions across the Galactic disk, a wavy pattern emerges in the variation of mean interstellar extinction as a function of Galactic longitude. 

 Although clusters at the same Galactic longitude show a range of reddening values due to their varying positions across the Galactic disk, a wavy pattern emerges in the variation of mean interstellar extinction as a function of Galactic longitude. Notably, extinction is relatively lower in the region of $l \sim 220^\circ$, while clusters in the region around $l \sim 40^\circ$ tend to have higher extinction. These variations reflect differences in interstellar material distribution, which align well with the observational data on inter-arm and spiral arm tangencies reported by \cite{2014A&A...569A.125H}. When we drew a best fit sinusoidal curve in Figure~\ref{fig04}, we obtained an equation of the form
\begin{equation}
\hspace{5.1cm} A_V = 2.15 + 0.81\sin(l + 49.1)
\end{equation}
\vspace{-0.5cm}
$$~~~~~\pm0.05 ~\pm 0.07~~~~~~\pm 4.9$$
This equation indicates that our view is most obscured by extinction in the direction of \(l = 41^{o} \pm 5^{o}\) and least affected in \(l = 221^{o} \pm 5^{o}\). A similar behavior was also observed by \citet{1998A&A...336..137C} and \citet{2005MNRAS.362.1259J}, although using a relatively smaller sample of Galactic OCs compared to the present dataset. Our findings are consistent with the recent results of \citet{2019A&A...625A.135L}, who observed wavy dust density patterns around the solar plane and noted that the solar neighborhood, within approximately 500 pc, remains atypical due to its extended structure above and below the Galactic plane.

A recent 3D reddening map based on Gaia, Pan-STARRS, and 2MASS data of almost half a billion stars by \citet{2019ApJ...887...93G} indicates that the color excess $E(g-r)$ along the line of sight in the direction of maximum extinction (\( l \sim 41^{o} \)) ranges from 0.58 to 3.36 mag, while in the direction of minimum extinction (\( l \sim 221^{o} \)) it ranges from 0.10 to 0.66 mag. These values correspond to visual extinction $A_V$ of approximately 2.61-13.36 mag towards \( l \sim 41^{o} \) and 0.45-2.97 mag towards \( l \sim 221^{o} \). Although our sinusoidal extinction variation is not directly comparable to a full 3D extinction map, we find that our estimated maximum and minimum extinctions  $A_V$(max) = 2.96 mag towards  \( l \sim 41^{o} \) and  $A_V$(min) = 1.34 mag towards \( l \sim 221^{o} \) are consistent with the values reported by \citet{2019ApJ...887...93G} at distances of roughly 1.2 kpc and 1.7 kpc, respectively. Given that the completeness of the cluster sample is expected to lie between 1 and 2 kpc, this alignment supports the reliability of our results. However, we caution that constructing a fully reliable and accurate 3D extinction model based solely on OCs remains difficult until a more extensive and distant cluster dataset becomes available.

%
%--------------------------- Fig. 05 -------------------------
\begin{figure*}
\centering
\includegraphics[scale=0.335]{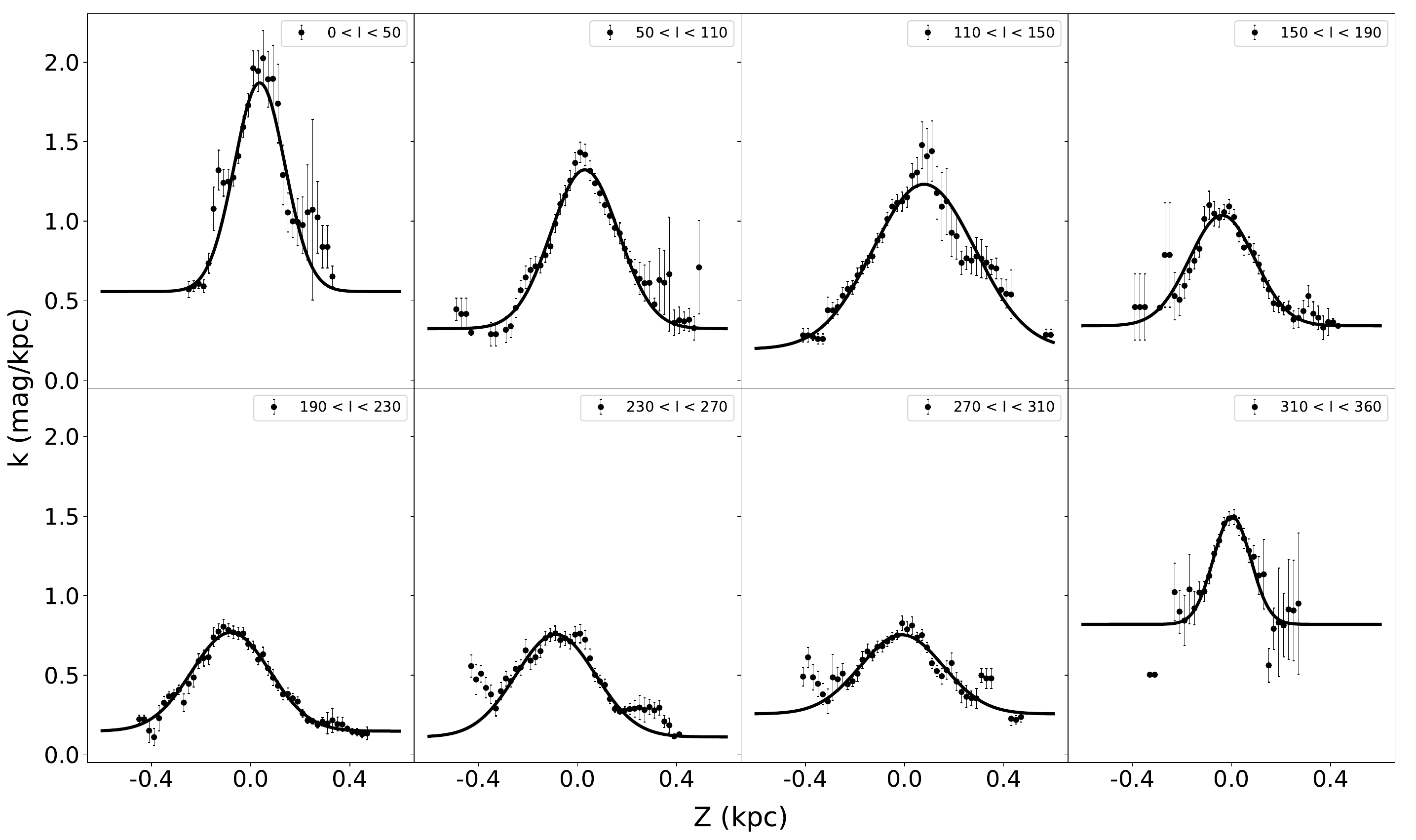}
\caption{The variation of $k$ as a function of $z$ for 8 different regions in longitude. The bin size in $z$ is 20 pc. Each point in the figure is weighted in proportion to the number of contributing clusters. A least square Gaussian fit in the distribution around the maxima is also shown with a thick continuous line and resultant mean values are given at the top of each panel.}
\label{fig05}
\end{figure*}
%----------------------------------------------------------------
%
\section{Galactic absorption and reddening plane}\label{galactic_absorption}
A key feature of interstellar matter in the Galaxy is its irregular structure, which complicates the process of mapping extinction as a function of Galactic longitude. Extinction increases with distance, and different longitude bins correspond to distinct distance ranges. However, this trend is less evident in high-latitude fields due to their low extinction and the likelihood that most stars are situated beyond the dust layer. In contrast, the trend is more pronounced in low-latitude fields \citep{2014MNRAS.443.1192C}, where the OCs used in this study are located. To account for the influence of distance, we computed the normalized interstellar absorption coefficient \( k \) for each cluster by dividing the extinction by the corresponding heliocentric distance \( d \).
\begin{equation}
\hspace{5.1cm} k = \frac{A_v}{d}
\end{equation}
Since all the selected clusters are confined to the thin disk, we did not normalize the sample for variations in  \( z \). To examine the concentration of reddening material in different directions around the Galactic plane, the sky was broadly divided into eight zones based on Galactic longitude. These zones are not uniform in size; instead, their boundaries were chosen so that regions near the Galactic center have larger bins, while those farther away have smaller bins. This approach accounts for the lower number of detected OCs per degree of longitude in the direction of the Galactic center due to higher extinction. Although this results in irregular boundaries, it helps reduce scatter within the selected bins. Figure~\ref{fig05} presents the distribution of interstellar absorption  \( k \) as a function of  \( z \) across eight distinct zones, offering a broad perspective on how absorption varies with vertical distance from the Galactic plane. A strong concentration of interstellar absorption within the Galactic plane is clearly evident, with significantly higher absorption in the direction of the Galactic center and relatively lower absorption toward the Galactic anti-center.
\input{table01.tex}
To analyze how the properties of the interstellar medium vary across different longitudinal directions in the solar neighborhood, a Gaussian profile was fitted around the maximum absorption in each zone. While other functions, such as exponential and  sech$^2$, were also tested, the Gaussian function was found to be the most suitable for describing the distribution around the peak. From the best-fit Gaussian profiles, we extracted various parameters, such as the maximum value of Galactic absorption $k_0$, the corresponding distance from the Galactic plane $z_0$, and half-width value $\beta$ of the reddening strip for all the zones. The parameters obtained from the best-fit Gaussian profiles are summarized in Table~\ref{tab:1}. The correlations of these parameters with Galactic longitude are examined in the following analysis.%
%-------------------------- Fig. 06 ------------------------------------
\begin{figure}[b]
\centering
\includegraphics[scale=0.60]{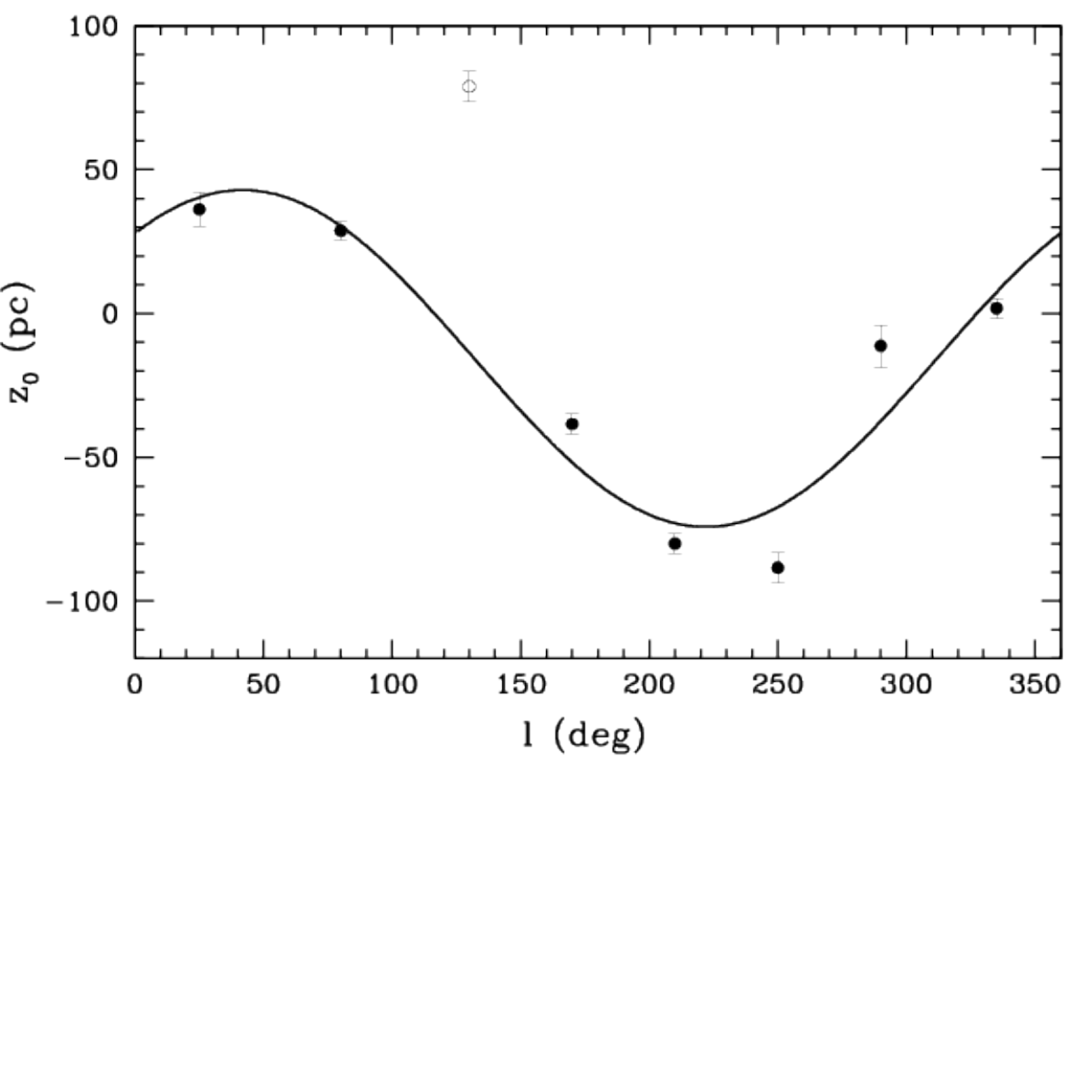}
\vspace{-3.5cm}
\caption{The height above or below the Galactic mid-plane $z_\odot$ at the maximum absorption $k_0$ is plotted as a function of Galactic longitude. The least square sinusoidal fit is shown by a continuous line. The point shown by an open circle is not included in the fit.}
\label{fig06}
\end{figure}
%--------------------------------------------------------------
%
%
\subsection{Variation in $z_0$ with longitude}\label{z0_l}
To study the variation of maximum absorption perpendicular to the Galactic plane, we plotted the value of $z_0$ against the mean value of longitude in Figure~\ref{fig06}. The height $z$ at the maximum absorption $k_0$ clearly shows a north–south asymmetry in the distribution. A least square solution for a sinusoidal function gives
\begin{equation}
\hspace{5.1cm} z_0 = -15.7 + 58.5\sin(l + 48.1)
\end{equation}
\vspace{-0.5cm}
$$~~~~~\pm 7.3 ~\pm 9.6~~~~~~~~~\pm 10.8$$
Here we have excluded one outlier point shown by the open circle from the least squares fit to ensure that the fitted function accurately represents the majority of the data. The best fit shows that the distance of the Galactic plane at maximum absorption is symmetric, with $z \sim -15.7\pm7.3$ pc. This indicates that a greater concentration of interstellar material lies below the formal Galactic mid-plane, in a plane commonly referred to as the reddening plane. The Sun is located $15.7\pm7.3$ pc above this reddening plane, a displacement often known as the solar offset (\(z_\odot\)). Accurately determining \(z_\odot\) is crucial for Galactic models, as it influences the line of sight when observing regions near the Galactic plane, thereby affecting the accurate estimation of the scale height.
%
%-------------------------- Fig. 07 ------------------------------------
\begin{figure}[b]
\centering
\includegraphics[scale=0.60]{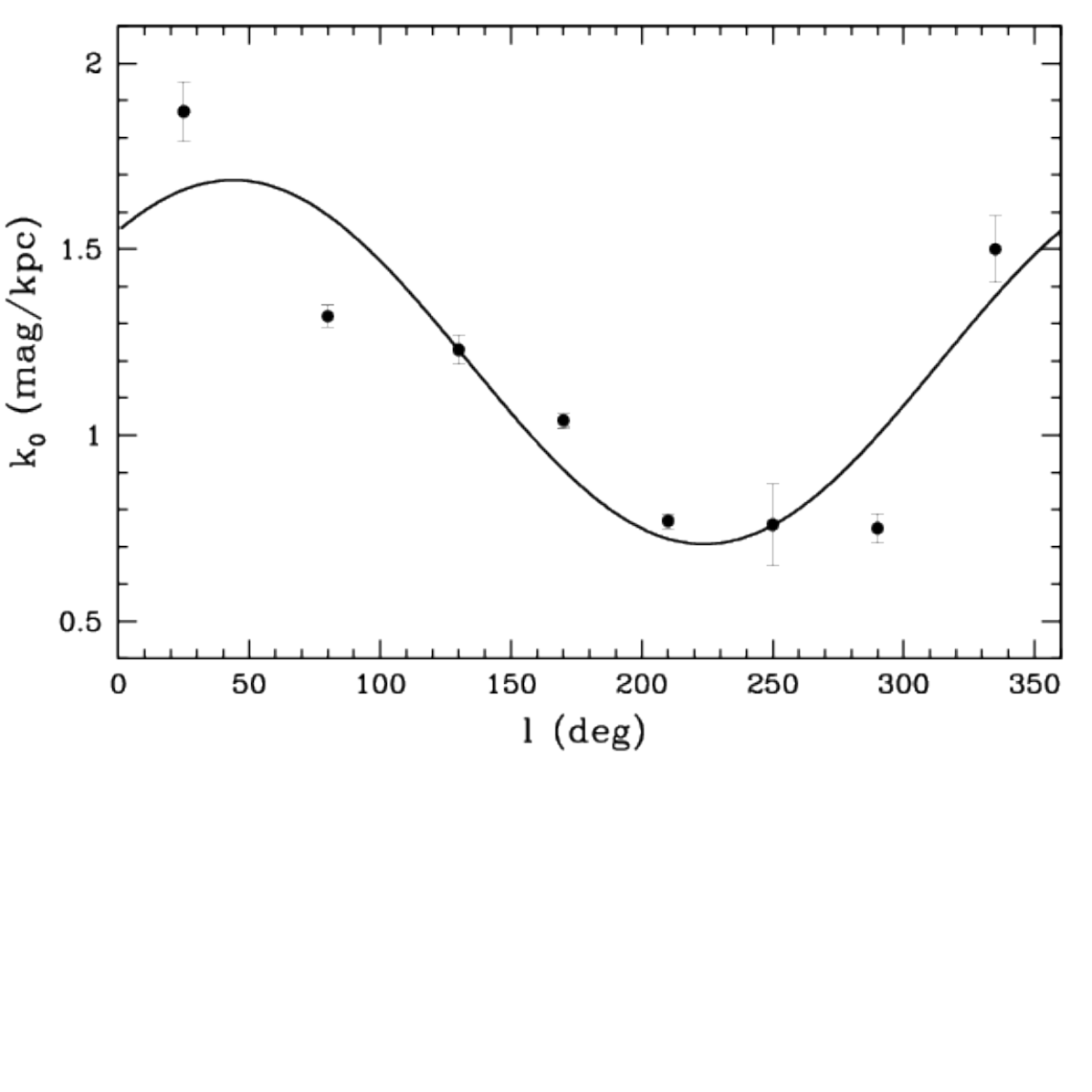}
\vspace{-3.5cm}
\caption{Maximum absorption $k_0$ as a function of Galactic longitude. A least square sinusoidal fit is drawn by a continuous line.}
\label{fig07}
\end{figure}
%----------------------------------------------------------------

Several studies have estimated $z_\odot$ using different stellar populations, including OB stars, OCs, pulsars, methanol masers, Wolf-Rayet stars, Cepheid variables, HII regions and giant molecular clouds. Most of these studies find $z_\odot$ to lie in the range of 10 to 25 pc north of the Galactic plane \citep[e.g.,][]{2007MNRAS.378..768J, 2009MNRAS.398..263M, 2014MNRAS.444..290B, 2017MNRAS.468.3289Y, 2019A&A...632L...1S, 2020AA...640A...1C, 2021MNRAS.502.4194G, 2023AJ....166..170J}. These variations are mostly attributed to the choice of data sample and the method used to determine \(z_\odot\), rather than to source selection \citep{2007MNRAS.378..768J, 2020AA...640A...1C}. The solar offset value determined based on the reddening plane is in excellent agreement with a similar estimate using the number density of clusters as derived in \citet{2023AJ....166..170J} as well as median solar offset derived by \citet{2017MNRAS.465..472K} using a wide range of stellar populations. Our estimated value of \(z_\odot\) is also well within the range of other such determinations. We also find that the direction of the maximum shift in the reddening plane, $\phi = 41.9^\circ$, lies within the first Galactic quadrant. A similar study by \citet{2006JRASC.100..146R}, which analyzed the maximum upward tilt of the plane defined by OB stars, also identified a peak near $l \sim 40^\circ$, consistent with our findings.
\subsection{Variation in $k_0$  with longitude}\label{abs_l}
As we have already observed, Galactic absorption reaches a minimum in the anti-center direction and a maximum towards the Galactic center, a region characterized by higher densities of gas and dust. To study this variation in greater detail, we present the variation in the maximum Galactic absorption, $k_0$, as a function of the Galactic longitude which is illustrated in Figure~\ref{fig07}. The absorption varies with longitude in a sinusoidal pattern, indicating a systematic variation in the amount of interstellar material along different lines of sight. The absorption is conspicuously low in the third Galactic quarter,  consistent with the findings of \citet{2008ApJ...672..930V}. A least squares solution to the absorption variation over the Galactic longitudes yields the following relation.   
\begin{equation}
\hspace{5.1cm} k_0 = 1.19 + 0.48\sin(l + 46.2)
\end{equation}
\vspace{-0.5cm}
$$\pm0.07 ~\pm 0.10~~~~~\pm 12.1$$
The absorption is found to be maximum towards $l \sim 44^\circ\pm 12^\circ$, and it is minimum in the direction of $l \sim 224^\circ\pm 15^\circ$. In our earlier study \citep{2005MNRAS.362.1259J} based on the 573 OCs within \(\lvert b \rvert \leq 5^{\circ}\), we also found that absorption approximately follows a sinusoidal curve with Galactic longitude, with the highest values occurring at $l \sim 30-50^\circ$ and the lowest values around $l \sim 220-250^\circ$. Extinction is generally low in the third Galactic quadrant, increasing only $\sim$ 0.07 mag kpc$^{-1}$ between $160^\circ < l < 280^\circ$. Except for the opaque barrier at $l \sim 265^\circ$ due to the Vela molecular ridge, the low-absorption window continues in the fourth Galactic quadrant towards Carina \citep[e.g.,][]{2010IAUS..266..106M}. The existence of a wide absorption window was also noticed by \citet{1968AJ.....73..983F} and \citet{2008ApJ...672..930V} in the region from $l \sim 215-255^\circ$ with a slightly enhanced value between $230^\circ$ to $240^\circ$, believed to be due to the presence of absorbing regions located between 1 and 2 kpc.
%
%-------------------------- Fig. 08 ------------------------------------
\begin{figure}[b]
\centering
\includegraphics[scale=0.60]{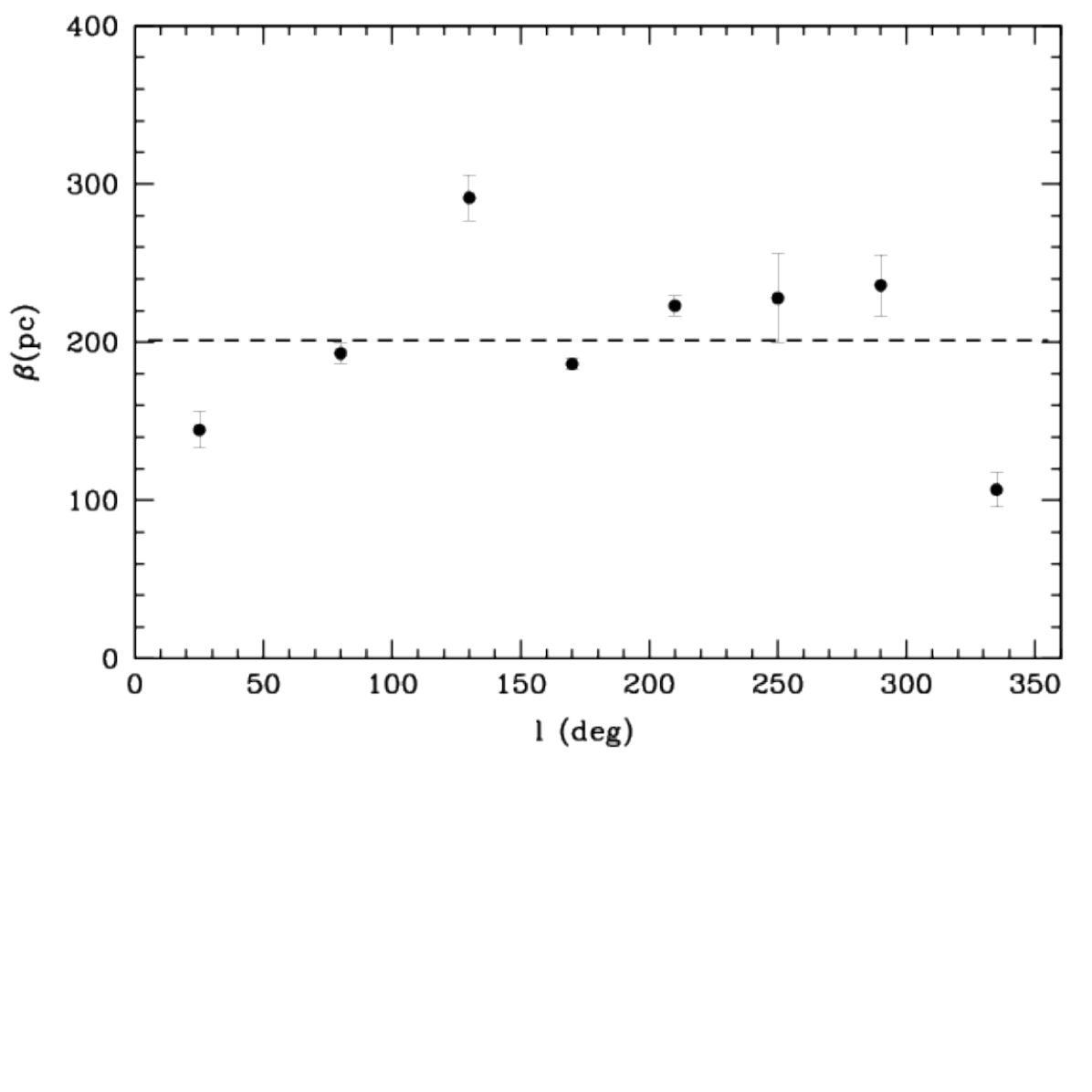}
\vspace{-3.5cm}
\caption{The half-width $\beta$ of the Gaussian distribution as a function of Galactic longitude. The mean half-width \(\beta\) is drawn by a continuous dashed line.}
\label{fig08}
\end{figure}
\subsection{Thickness of the reddening plane}\label{rp_thick}
The thickness of the reddening plane is a measure of the extent or depth of dust layer in the interstellar medium (ISM), which affects the amount and distribution of light absorption and scattering across different wavelengths. It is defined by the half-width value, $\beta$, of the Gaussian distribution, which is estimated as half of the separation between the $z$ values at which the interstellar absorption drops to $1/e$ times of its maximum value \citep{1987MNRAS.226..635P}. The value of $\beta$ is not uniform across the Galaxy and can vary depending on the location within the Galactic disk. In Figure~\ref{fig08}, we illustrate variation of $\beta$, as estimated for each zone, as a function of mean Galactic longitude. It is evident that $\beta$ varies with Galactic longitude, ranging from approximately 107 pc to 291 pc in different directions with a mean thickness of $\beta = 201\pm20$ pc. The thickness of the dust layer is not uniform and can vary significantly depending on local conditions such as the presence of molecular clouds and star-forming regions.

The distribution of reddening material within a few kiloparsecs of the solar neighborhood has also been studied by \citet{1968AJ.....73..983F} and \citet{1987MNRAS.226..635P}. While the former found \(\beta\) to vary between 40 to 100 pc, the latter obtained values ranging from 50 to 210 pc. In our earlier study in \citet{2005MNRAS.362.1259J} using 573 OCs within \(\lvert b \rvert \leq 5^{\circ}\), we obtained the mean value of \(\beta\) as 125 \(\pm\) 21 pc. These differences suggest varying estimates for the thickness of the reddening layer across different studies, however, all three previous studies were based on smaller samples of OCs.
%
%--------------------------------------------------------------
%
%--------------------------- Fig. 09 -----------------------------------
\begin{figure}[b]
\centering
\includegraphics[scale=0.35]{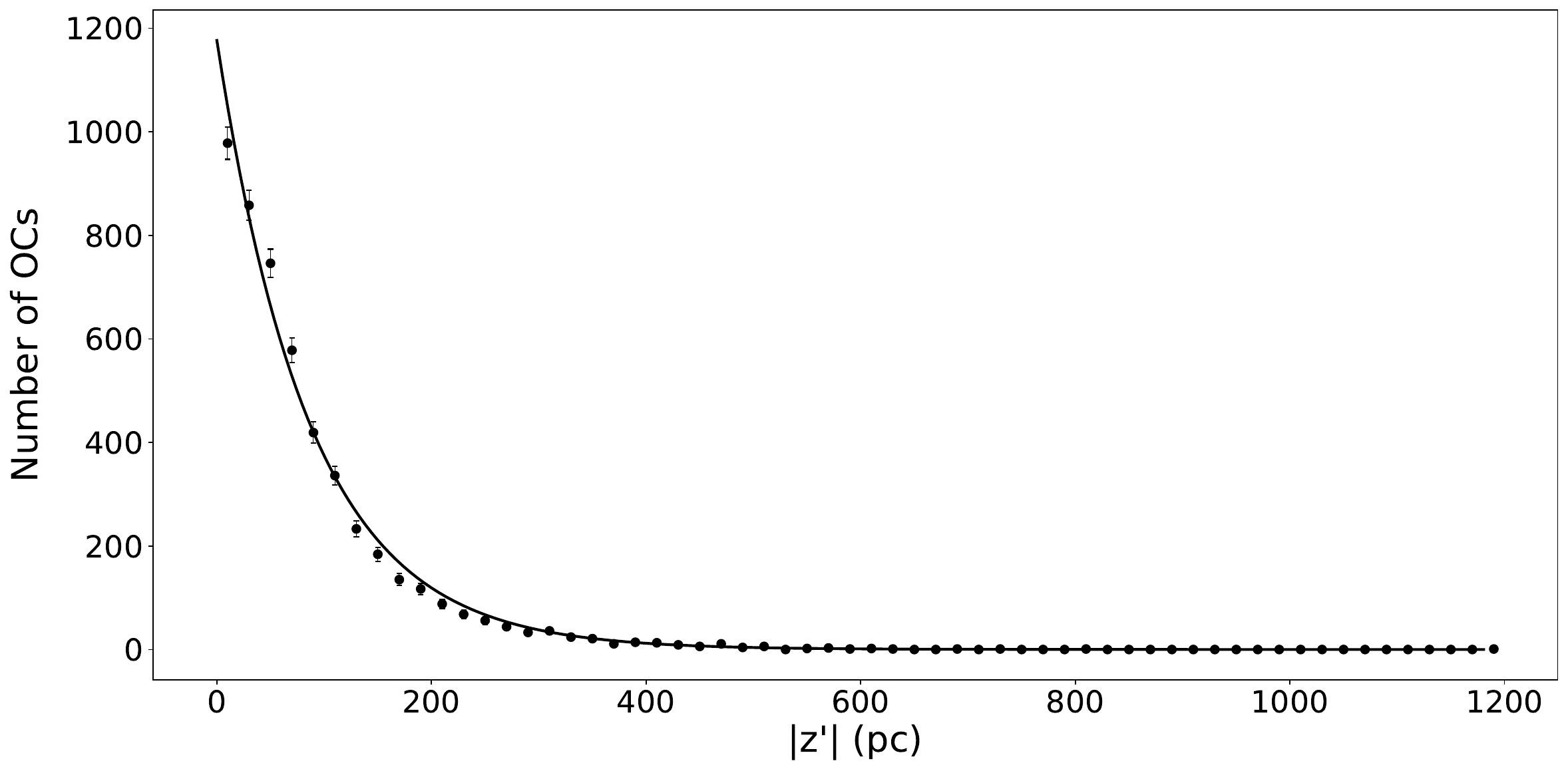}
\caption{\(\lvert z' \rvert\) distribution of OCs with a bin size of 20 pc. The least square exponential decay profile is shown by a continuous line. The error bars shown in the y-axis are the Poisson errors.}
\label{fig09}
\end{figure}
%-------------------------------------------------------------
%
\section{Cluster scale height from the reddening plane}\label{scale_height}
The vertical number density of OCs in the Galactic disk decreases exponentially with distance from the Galactic mid-plane. The exponential decrease is characterized by a decay factor, known as the scale height, which parametrizes the thickness of the Galactic disk \citep{2025A&A...694A..70M}. In a similar manner, the thickness of the reddening plane can be determined by obtaining its scale height from the reddening plane \citep{2005MNRAS.362.1259J}. Since reddening plane is slightly inclined with respect to the formal Galactic plane, we used the following relation defined by \citet{1968AJ.....73..995F} to determine the vertical distance of a cluster from the reddening plane,
\begin{equation}
\hspace{5.1cm} 
z' = z_\odot + d \sin b \cos \phi - d \cos b \sin \phi \cos (l - \theta_t)
\end{equation}
where \(z_\odot\) is the solar offset, $l$ is the Galactic longitude, \(\phi\) is the inclination angle, and \(\theta_t\) is the angle of maximum inclination of the reddening plane with respect to the formal Galactic plane. We took \(\theta_t\) = \(48^\circ\pm 7^\circ\), and $z_\odot$ as $-15.7\pm7.3$ as determined in the previous section. We considered \(\phi = 0.25^\circ \pm 0.16^\circ\) as given in \citet{1987MNRAS.226..635P}. In Figure~\ref{fig09}, we draw histogram of the $z^\prime$ distribution. The number density distribution of $z^\prime$ is described by the decaying exponential disk of the form,
\begin{equation}
\hspace{5.1cm} 
N(z) = N_0~exp \left(- \frac{|z'-z_0|}{z_h}\right). 
\end{equation}
where $z_h$ is the scale height of the distribution characterized as thickness of the reddening plane. Since our sample of OCs is confined within $|b| \leq 6^0$, the above distribution broadly represents the thin disk component, where most of the interstellar material resides. We fit the exponential profile to the number distribution of OCs to estimate the scale height, as illustrated in Figure~\ref{fig09}. The scale height of the distribution of clusters in the reddening plane is found to be $z'_h = 87.3\pm1.8$ pc. However, considering only young clusters having ages less than 700 Myr, we find a smaller scale height of $z'_h = 80.8\pm2.8$ pc for the clusters distribution around the reddening plane. This is consistent with the well-established notion that scale height increases with the age of the clusters \citep{2016A&A...593A.116J, 2020A&A...640A...1C}. This variation in scale height with age is interpreted as evidence either for a thicker disk in the past or for dynamical heating of the disk due to scattering events, such as interactions with spiral arms and giant molecular clouds, or stronger tidal disruption near the Galactic plane  \citep{2014MNRAS.444..290B, 2025A&A...694A..70M}.

Several estimates of the scale height of the reddening material exist, but these vary depending on the wavelength of observation, the region of the disk studied, and the methodology used. While \citet{1982A&A...109..213L} found a scale height of 130 pc, \citet{1987MNRAS.226..635P} reported the value of \(160\pm20\) pc for the distribution of reddening material. From the Two Micron All Sky Survey data, \citet{2006A&A...453..635M} obtained a scale height of the interstellar matter as $125^{+17}_{-7}$. pc. While most of these studies found the scale height in between 120 to 160 pc, the present estimate of $z'_h = 87.3\pm1.8$ pc is substantially smaller than earlier results. However, such a discrepancy is not unexpected, given the differing sample sizes, incompleteness effects, and uncertainties in extinction measurements in earlier studies. On the other hand, from a compilation of different CO surveys, \citet{1998A&A...330..910M} found $z'_h$ in between 40 to 70 pc which is closer to the present estimate. It is however evident from this study that the observed scale height of the matter distribution in the stellar disk is consistent with the vertical number density distribution of the clusters.
\section{Summary and Conclusions}\label{summary}
The interstellar material, composed of gas and dust, is distributed unevenly throughout the Galactic disk, forming a complex structure influenced by Galactic dynamics. Mapping the distribution of this material in the Galactic disk is essential for understanding the composition and evolution of the Milky Way, as well as for tracing the Galactic structure. The primary goal of this study was to perform a comprehensive analysis of the distribution of reddening material at low Galactic latitudes based on reddening and interstellar extinction data available for the largest sample of open clusters to date. To achieve this, we utilized various recently released catalogs of Galactic open clusters and compiled a total of 6,215 open clusters with available reddening and distance information. This sample significantly surpasses the 722 open clusters used in the last such study by \citet{2005MNRAS.362.1259J}.  It is evident from the distribution that more than 80\% of the clusters lie within $-6^\circ \le b \le 6^\circ$ of the Galactic mid-plane, where most of the reddening material lies.

We constructed a 2D extinction map of the Galactic disk and found a strong concentration of absorbing material near the Galactic mid-plane, with a gradual decrease in vertical direction from the Galactic mid-plane. The slope of $\frac{dA_V}{dz}$ is found to be approximately $0.9\pm0.1$ mag pc$^{-1}$, confirming a systematic decline in interstellar matter with increasing altitude. Interstellar absorption is observed to vary rapidly with small changes in Galactic longitude. We found that the absorption follows a sinusoidal variation with Galactic longitude, indicating systematic differences along various lines of sight. It also varies sinusoidally both above and below the Galactic mid-plane. Absorption is at a minimum in the anti-center direction and reaches a maximum towards the Galactic center, known for its dense regions of gas and dust. The maximum and minimum interstellar extinction occur near $\sim 41^\circ\pm 5^\circ$ and $\sim 221^\circ\pm 5^\circ$, respectively.

We further determined the interstellar absorption $k$ by normalizing the interstellar extinction by the corresponding heliocentric distance $d$ for each cluster. The absorption shows asymmetry around the Galactic plane, suggesting the presence of a reddening plane slightly skewed below the formal Galactic plane. This offset is found to be about $15.7\pm7.3$ pc above the Galactic plane of symmetry defined by the reddening material and is believed to represent the current value of the solar offset. This result is consistent with our previous estimate based on the number density of clusters, as well as numerous other studies. Additionally, we observed that the reddening plane is not strictly co-planar with the formal Galactic plane but exhibits a wavy pattern around the Galactic mid-plane. The maximum and minimum height of the dust plane occurring at $\sim 42^\circ$ and $\sim 222^\circ$, respectively which coincide with the maximum and minimum absorption found at $\sim 44^\circ$ and $\sim 224^\circ$, respectively. The reddening plane is found to be thin compared to the broader Galactic disk and the mean thickness of the absorbing material in the reddening plane, as determined by the half-width value $\beta$, is estimated to be about $201\pm20$ pc though with substantial variations in different longitudinal directions. The thickness of the reddening plane has important implications for understanding the Galactic structure and the distribution of interstellar matter. Analyzing its variation across different regions of the Galaxy provides valuable insights into the physical processes driving the evolution of the interstellar medium, the effects of stellar feedback, and the dynamical behavior of the Galactic disk over time.

We determined a scale height of $87.3\pm 1.8$ pc for the distribution of reddening material which is slightly smaller than the distribution of OCs around the Galactic mid-plane as obtained in \cite{2023AJ....166..170J} but within broad range of 50-100 pc given in most of the studies. Overall, this study offers valuable insights into the non-uniform distribution of interstellar dust in the Galactic disk and enhances our understanding of the complex interplay between open clusters in the thin disk and the surrounding interstellar medium in the reddening plane. A major limitation in analyzing the reddening plane is the lack of a significant number of open clusters beyond 3 kpc. Future multi-band observations of distant open clusters, particularly with the upcoming Gaia DR4 release in 2026, are expected to significantly improve our understanding of the connection between interstellar material and Galactic disk structure by enabling deeper and more accurate mapping across extended Galactocentric distances. Furthermore, incorporating data from upcoming surveys such as the Vera C. Rubin Observatory’s Legacy Survey of Space and Time (LSST) will allow for a more precise study of the dust structure and its variation across the Galactic disk.
\section*{Acknowledgments}
I would like to thank Sagar Malhotra for helping me out in the data analysis as part of his master project. I am grateful to Jayanand Maurya for stimulating discussions over the course  of this work. This work has made use of data from the European Space Agency (ESA) mission Gaia (https://www.cosmos.esa.int/gaia), processed by the Gaia Data Processing and Analysis Consortium (DPAC, https://www.cosmos.esa.int/web/gaia/dpac/consortium). 
\bibliography{manuscript.bib}

\end{document}

%% file: table01.tex
\begin{table*}
%  \centering
  \caption{Parameters obtained from the Gaussian fits in the 8 longitudinal zones. Here, $k_0$, $z_0$, and $\beta$ represent maximum value of absorption, corresponding distance from the Galactic plane, and thickness of the absorbing material as derived from the best fit Gaussian profiles.}
    \begin{tabular}{ccccrrrr}
    \hline
    Longitude range & Number of & {$k_0$}  & {\(\sigma(k_0)\)}  & {$z_0$} & {\(\sigma(z_0)\)}  & {$\beta$}  & {\(\sigma(\beta)\)} \\
    (deg)       & open clusters& (mag/kpc) & (mag/kpc) & (pc) & (pc) & (pc) & (pc)   \\
    \hline  \\         
     0 - 50   & 495   & 1.87  & 0.08  &  36.2  & 5.9   & 144.6   & 11.4 \\
    50 - 110  & 891   & 1.32  & 0.03  &  28.7  & 3.2   & 193.0   &  6.2 \\
    110 - 150 & 524   & 1.23  & 0.04  &  78.9  & 5.3   & 291.3   & 14.3 \\
    150 - 190 & 401   & 1.04  & 0.02  & -38.4  & 3.5   & 186.2   &  3.8 \\
    190 - 230 & 547   & 0.77  & 0.02  & -80.0  & 3.3   & 223.1   &  6.0 \\
    230 - 270 & 713   & 0.76  & 0.11  & -88.4  & 5.2   & 227.9   & 28.0 \\
    270 - 310 & 781   & 0.75  & 0.04  & -11.3  & 7.4   & 236.0   & 19.5 \\
    310 - 360 & 689   & 1.50  & 0.09  &   1.8  & 3.5   & 106.8   & 10.9 \\
    \hline
    \end{tabular}%
  \label{tab:1}%
\end{table*}%